\documentstyle[12pt,epsf,amstex]{article}
\begin{document}
\setlength{\unitlength}{1mm}
\textwidth 15.0 true cm %\sect
\headheight 0 cm
\headsep 0 cm
\topmargin 0.4 true in
\oddsidemargin 0.25 true in
\input epsf

\newcommand{\beq}{\begin{equation}}
\newcommand{\eeq}{\end{equation}}
\newcommand{\be}{\begin{eqnarray}}
\newcommand{\ee}{\end{eqnarray}}
\renewcommand{\vec}[1]{{\bf #1}}
\newcommand{\vecg}[1]{\mbox{\boldmath $#1$}}
\renewcommand{\theequation}{\thesection.\arabic{equation}}
\newcommand{\grpicture}[1]
{
    \begin{center}
        \epsfxsize=200pt
        \epsfysize=0pt
        \vspace{-5mm}
        \parbox{\epsfxsize}{\epsffile{#1.eps}}
        \vspace{5mm}
    \end{center}
}

\begin{flushright}

SUBATECH--2002/05\\
ITEP-TH--03/02\\

\end{flushright}

\vspace{0.5cm}

\begin{center}

{\Large\bf  On the relation between effective supersymmetric actions

in different dimensions.}

\vspace{1cm}

  {\Large  E.T. Akhmedov} \\

\vspace{0.5cm}

{\it 117259, ul. B.Cheremushkinskaya, 25, ITEP, Moscow} \\
\vspace{1cm}
and

\vspace{1cm}

{\Large A.V. Smilga} \\

\vspace{0.5cm}

{\it SUBATECH, Universit\'e de
Nantes,  4 rue Alfred Kastler, BP 20722, Nantes  44307, France. }\\

\end{center}

\bigskip

\begin{abstract}
We make two remarks: {\it (i)} Renormalization of the effective
charge in a 4--dimensional (supersymmetric) gauge theory is
determined by the same graphs and is rigidly connected to the
renormalization of the metric on the moduli space of the classical
vacua of the
corresponding reduced quantum mechanical system. Supersymmetry
provides constraints
for possible modifications of the metric, and this gives us a
simple proof of nonrenormalization theorems for the original
4-dimensional theory.  {\it (ii)} We establish a nontrivial
relationship  between
the effective $(0+1)$--dimensional and  $(1+1)$--dimensional
 Lagrangia. (The latter represent conventional  K\"ahlerian
$\sigma$ models.)
\end{abstract}

\section{Introduction.}
Consider  4--dimensional supersymmetric gauge theory placed
in a small spatial torus $T^3$ of size $L$.
We assume that
$g^2(L) \ll 1$ and perturbation theory makes sense.
For unitary and symplectic gauge groups $G$, the only classical
vacua of this theory are given by constant gauge potentials
$A_k , \ k = 1,2,3$,
lying in the Cartan subalgebra of the group\footnote{It is not the case
for higher orthogonal and exceptional groups \cite{newvac}, but these
complications are beyond the scope of the present paper.} $G$.
 The low--energy dynamics of
the model is determined by the effective Hamiltonian describing
motion over the vacuum moduli space. Due to supersymmetry, the
energy of a classical vacuum configuration
stays zero after loop corrections are taken into account -- no
potential is generated.
\footnote{This is true only for nonchiral theories, which
are only considered in the present paper. In the theories
with chiral matter content, the situation is more complicated
\cite{jachir}.}
 However, supersymmetry
usually allows the existence of a nontrivial metric on the moduli
space, in which case such a metric is generated after loop
corrections are taken into account.

The  loop corrections to the effective Hamiltonian were
calculated first in Ref.\cite{jaSQED} in the simplest case of
${\cal N} = 1$ supersymmetric QED with two chiral matter multiplets
 of opposite charges. In this case, the moduli space is represented
by the constant gauge potentials $A_k$ and their
superpartners. Note
that, for a field theory on $T^3$, the moduli space is compact,
$0 \leq A_k \leq 2\pi/L$.

The original calculation was carried out
in the Hamiltonian framework. The effective Hamiltonian
is expressed in terms of $A_k$, $P_k = -i\partial/\partial\, A_k$,
and the zero Fourier mode
of the photino field $\psi_\alpha, \ \alpha = 1,2$. It has
the form
 \be
\label{Heff}
\frac 1{e^2}  H^{\rm eff} \ =
 \frac 12 f(\vec{A})   P_k^2 f(\vec{A}) - \nonumber \\
 - \epsilon_{jkp} \  {\bar\psi} \sigma_j \psi  f(\vec{A})
\partial_p f(\vec{A})  P_k
- \frac 12  f(\vec{A}) \partial^2_k f(\vec{A})
(  {\bar\psi} \psi )^2 \ ,
  \ee
where  $\sigma_i$ are the Pauli matrices and
 \be
\label{fQED}
f(\vec{A}) = 1 - \frac{e^2}{4} \, \sum_{\vec{n}}
\frac{1}{|\vec{A} L -
2  \pi \vec{n}|^3} +  \dots
 \ee
(we have rescaled $A \to A/e$ compared with
 the normalization of Ref. \cite{jaSQED}).
The dots stand for possible higher--loop corrections. The
expression (\ref{fQED}) is written for the theory where
the charged fields are massless.

Note that the sum in the right side of Eq. (\ref{fQED})
diverges logarithmically at large $|\vec{n}|$. This is none
other than the effective charge renormalization
\be
\label{runcharg}
e^2(L) \ =\ e_0^2 \left[1 - \frac {e_0^2}{4\pi^2}
\ln(\Lambda_{UV}L)  + \ldots \right]\ .
 \ee
In the massive case and if the box is large enough,
$\ln(\Lambda L)$ is substituted by $\ln(\Lambda/m)$.
On the other hand, if  we are dealing with
dimensionally reduced SQED, where
all Fourier harmonics with $\vec{n} \neq 0$ are ignored, we
 obtain
 \be
\label{fQEDQM}
f(\vec{A}) = 1 - \frac{e^2}{4 |L \, {\vec{A}}|^3}\ .
 \ee
It is obvious that the coefficients in Eq. (\ref{fQEDQM}) and
in Eq. (\ref{runcharg}) are related\footnote{The procedure for
getting the aforementioned relation is similar
to the T-duality transformation on D-branes in string theory
(see in particular Ref. \cite{Tdual}). In fact, $N=2$ SQED with
doublet of hypermultiplets (plus free uncharged hypermultiplets)
reduced to one dimension
can be interpreted as  the theory of a D0-brane in the vicinity
of a D4-brane. The latter system was extensively studied \cite{Tdual}.
 In this note we concentrate, however, on the
$N=1, d=4$ systems,
which were not considered so far in the string theory
framework.}. The knowledge of
$\beta$--function allows one to determine the modification
of the metric on the moduli space, and this is how the
effective Hamiltonian for ${\cal N} = 1$ non-Abelian theories
was evaluated in recent Ref. \cite{nonab}.
The inverse is also true, however, and this is one of the main
emphasize of the present paper. We note that
the $\beta$--function of field theories can be conveniently
calculated via modification of the metric
in the quantum mechanical limit where all nonzero Fourier harmonics
are ignored. Ideologically, this is
the ultraviolet cutoff procedure brought to its extreme.
One can call it ultraviolet {\it chopoff}.

As was mentioned, the result (\ref{Heff}) was first obtained
in the Hamiltonian framework using a systematic Born--Oppenheimer
expansion for $H^{\rm eff}$. There is, however, a simpler way
to derive the same result: to evaluate the term $\propto
\dot{\vec{A}} \dot{\vec{A}}$ in
 the effective
{\it Lagrangian} in a slowly varying bosonic background
$\vec{A}(t), \psi_\alpha = 0$. Other structures in the Lagrangian
can be restored using supersymmetry.
The plan of the paper is the following. In Sect. 2 we present
one--loop calculations of the effective Lagrangian. We use
the background field method and demonstrate that the result
is given by exactly the same graphs as the graphs determining
the 4--dimensional $\beta$--function. Like in 4--dimensional case,
the contribution of scalar deteminant cancels out in supersymmetric
case, and we are left with the graphs describing fermion and gauge
boson magnetic interactions.

Next, we go beyond one loop and prove in Sect. 3
non-renormalization theorems for four-dimensional
${\cal N} = 2$ and ${\cal N} = 4$
SYM theories. Sect. 4 is not devoted to the $\beta$ function, but addresses
a related question of the connections between effective models
in different dimensions. The Hamiltonian (\ref{Heff}) represents
a nonstandard $\sigma$ model. The model enjoys
${\cal N} = 2$ QM symmetry (it has 2 different complex
supercharges), but is not a K\"ahlerian model (K\"ahlerian
$\sigma$ models are defined on even--dimensional target
spaces, whereas the model (\ref{Heff})
 is defined on a 3--dimensional conformally flat manifold). We show
that the model (\ref{Heff}) is related, however, to
K\"ahlerian models in a nontrivial way: to obtain a K\"ahlerian
model out of Eq. (\ref{Heff}), one has to go back
to the original 4--dimensional field theory and consider it
on an asymmetric torus, with one of the sizes much larger than
the others.

\section{Ultraviolet chopoff and $\beta$ function.}

Let us consider for definiteness ${\cal N}=1$ four--dimensional
$SU(2)$ SYM theory
\begin{equation}
{\cal L} = \frac{1}{g^2}\,
{\rm Tr} \, \left( -\frac{1}{2} F_{\mu\nu}^2 +
{\rm i} \, \bar{\lambda} /\!\!\!\!{D} \lambda \right) \label{SYM}\ ,
\end{equation}
where  $\lambda$ are
Majorana fermions in the adjoint representation of $SU(2)$ and
$D_\mu = \partial_\mu - {\rm i} \left[A_\mu, \, \cdot
\right]$.
 Put the system in a small spatial box and impose the periodic boundary
 conditions.
 We would like to calculate quantum corrections to the effetive
action in the abelian background field $A_\mu = C_\mu t^3$,
$C_\mu = (0, C_i)$.  We assume that
$C_i$ varies slowly with time, but does not depend on spatial
coordinates\footnote{The setup of the problem is basically the same as in
Refs.\cite{sestry},\cite{DoPoStAk}. There are two differences: {\it (i)} 
We are
considering the ${\cal N} = 1$ rather than ${\cal N} = 4$ theory
and the corrections are not going to vanish. {\it (ii)} The authors
 of Ref. \cite{sestry} did their calculation bearing in mind the
geometric picture
of scattered $D0$ branes and their background was slightly more
sophisticated than ours.}. The background
fermionic fields (superpartners of $C_\mu$) are taken to be zero
at this stage.

The calculation can be conveniently done using background gauge
method \cite{background}. We decompose the field in the classical
(abelian) background and quantum fluctuations,
  \be
 \label{decompos}
A_\mu \ \to \ C_\mu t^3 + {\cal A}_\mu \ ,
 \ee

and add to the Lagrangian the gauge-fixing term
\be
\label{gaugfix}
- \frac 1{2g^2} (D_\mu^{\rm cl} {\cal A}_\mu)^2\ ,
\ee
where $D_\mu^{\rm cl} = \partial_\mu - {\rm i}
 \left[A_\mu^{\rm cl}, \, \cdot \right]$. In what follows
we use the notation $A_\mu \equiv A_\mu^{\rm cl}
= C_\mu t^3$. The coefficient chosen in Eq. (\ref{gaugfix}) defines
the ``Feynman background gauge'', which is simpler and more
convenient than others.
Adding (\ref{gaugfix}) to the first term in Eq. (\ref{SYM}) and
integrating by parts, we obtain for
 the gauge--field--dependent part of the Lagrangian
\begin{equation}
{\cal L}_{\cal A} =
-\frac{1}{2 g^2}\, {\rm Tr} \, \left(F_{\mu\nu}^2 \right) +
 \frac{1}{g^2} \, {\rm Tr}\,
\left\{ {\cal A}_\mu \left( D^2 g_{\mu\nu}{\cal A}_\nu  - 2i
\left[F_{\mu\nu}, {\cal A}_\nu  \right] \right)
 \right\} + \ldots \ ,
\end{equation}
where the dots stand for the terms of higher order in
${\cal A}_\mu$.
The ghost part of the Lagrangian is
\begin{equation}
{\cal L}_{ghost} = - 2\, {\rm Tr} \,\left( \bar{c} D^2 c \right)
\ + {\rm higher\ order\ terms}
\end{equation}

Now we can integrate   over the quantum fields ${\cal A}_\mu$,
$c$, and
also over the fermions using the relation
$$
\left({\rm i}\,/\!\!\!\!{D}\right)^2 = - D^2 +
\frac{\rm i}{2} \sigma_{\mu\nu} F_{\mu\nu}\ ,
$$
$\sigma_{\mu\nu} = \frac 12 [\gamma_\mu, \gamma_\nu]$.
We obtain the effective action as follows:
   \be
\label{detratio}
S_{\rm eff} = -\frac{1}{2 \, g^2} \, \int_{T^3\times R}
{\rm Tr} (F_{\mu\nu}^2)\ + \nonumber \\
\log  \left(\frac{\det^{\frac14}\left(- D^2 \, I + \frac i2
\sigma_{\mu\nu}  \left[F_{\mu\nu}, \, \cdot\right]\, \right)
\det\left(- D^2\right)}{\det^{\frac12}\left(-D^2\, g_{\mu\nu} +
2i\, \left[F_{\mu\nu}, \, \cdot\right]\right)}\right)\ .
   \ee
We see that the fermion and gauge field determinants involve,
besides the
term $-D^2$ which is present also in the scalar determinant, the
term
$\propto F_{\mu\nu}$ describing the magnetic moment interactions. An
important
observation is that, were these magnetic interactions absent, the
contributions
of the ghosts, fermions, and gauge bosons would just cancel
 and the effective action would not
acquire any corrections. This feature is common for all
supersymmetric
gauge theories (${\cal N} = 1$, ${\cal N} = 2$, ${\cal N} = 4$;
non-Abelian and Abelian). This fact is related to another known fact
that, when supersymmetric $\beta$
function is calculated in the {\it instanton}
background, only the contribution of the zero modes survives
\cite{shif}.

For nonsupersymmetric theories,
also nonzero instanton modes provide  a nonvanishing constribution
in the
 $\beta$ function. On the other hand, the contributions due to
det $(-D^2)$ in the effective action do not vanish in the
nonsupersymmetric case.

\begin{figure}
   \begin{center}
        \epsfxsize=300pt
        \epsfysize=100pt
        \vspace{-5mm}
        \parbox{\epsfxsize}{\epsffile{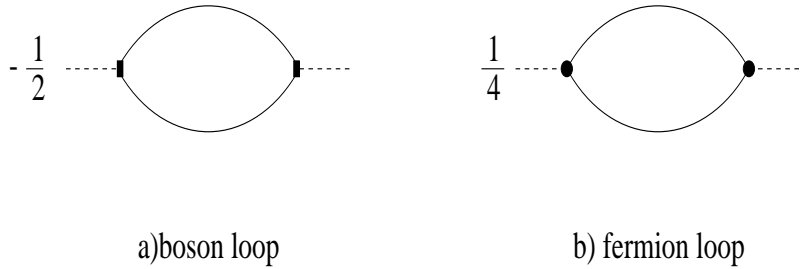}}
        \vspace{-5mm}
    \end{center}
\caption{One-loop renormalization of the kinetic term in SYM. Internal lines
are Green's functions of the operator $(-D^2)$. The vertices involve the spin
operator $J_{\alpha\beta}$ 
and are different for the fermion and gauge boson loop.}
\label{dvepetli}
\end{figure}

To find the magnetic contributions, we have to calculate the graphs
 drawn
in Fig. 1. The vertices there are proportional to $\epsilon^{abc}
F_{\alpha\beta} J_{\alpha\beta}$ ($J_{\alpha\beta}$ being the spin
operator
in the corresponding representation) and the lines are  Green's
functions
of the operator $-D^2$.
Only the color components 1 and 2 circulate in the loops. They
acquire the
mass $|\vec{C}|$ in the Abelian background $C_i t^3$.
One can be convinced that gauge boson loop involves
the factor $-4$ compared to the fermion one [the factor
$\frac12: -\frac 14
= -2$ is displayed in Eq. (\ref{detratio}) and Fig. 1 and another factor 2
comes from  spin;
see Eq. (16.128) in Peskin's book].

Let us calculate, say, the fermion loop. If all higher Fourier modes
are ``chopped off'', $-D^2 \to  -\partial_0^2 - \vec{C}^2$ and
the corresponding contribution to the effective Lagrangian
is
  \be
 \label{feract}
 \frac14 \cdot \frac12 \cdot
2 \cdot \dot C_j \dot C_k \, {\rm Tr} \{ \sigma_{0j}\sigma_{0k} \}
\int_{-\infty}^\infty \frac{d\omega}{2\pi} \frac 1{(\omega^2 +
\vec{C}^2)^2}
\ = \  \frac {{\dot \vec{C}}^2}{4|\vec{C}|^3}  \ .
 \ee

(the factor $\frac14$ is the power of the determinant in
Eq. (\ref{detratio}),
$\frac12$ comes from the expansion of the logarithm and 2 is the
color
factor.)
Adding the gauge boson contribution and also the free bosonic term,
 we obtain
  \begin{equation}
  \label{Leffbos}
  \frac {g^2}{L^3} {\cal L}^{\rm eff}_{\rm bos}\ = \
  \frac{ \dot{\vec{C}}^2}{2 f^2(\vec{C})} \,
  \end{equation}
with
 \be
\label{fnab}
 f(\vec{C}) \ =\ 1 + \frac {3g^2}{4L^3 |\vec{C}|^3} + \ldots \ .
 \ee
If higher Fourier modes are taken into account, we obtain in the
exact
analogy with Eq. (\ref{fQED})
\be
 \label{fasym}
f(\vec{C}) = 1 +  \frac{3 g^2}{4} \, \sum_{n_k}
\frac{1}{\left[\sum_k (C_k \, L_k -
2\, \pi \, n_k)^2 \right]^{3/2}}
\ + \ldots \, \ ,
\ee
where, bearing in mind further applications, we assumed that the
sizes
of the torus  $L_k$, $k=1,2,3$, do not coincide.
The sum is divergent at large $n_k$. The coefficient of the
logarithm  gives the $\beta$ function of
the ${\cal N} = 1$ SYM theory. For sure, this could be expected in
advance.
What is not quite trivial, however, is that the calculation in the
truncated
theory is {\it absolutely} parallel to the well-known calculation
in 4 dimensions \cite{background}: in four dimensions the
corrections to the effective action are  also given by the
graphs in Fig. 1, and
the gauge boson and the fermion contributions in  the
$\beta$ function
have the respective coefficients\footnote{A remarkable fact is that
one obtains the same ratio calculating the effective action
in the instanton background field: the correct coefficient 6 in the
$\beta$ function is obtained as 8 - 2, where ``8'' is the number
of bosonic zero modes
and ``2''  is a half of the number of fermionic zero modes in the
instanton background.} $4:-1$.
  We will see soon that this is specific for
 supersymmetric theories. In nonsupersymmetric case, the $\beta$
function can
also be calculated with the chopoff technique, but the relevant
graphs
are different.

  The bosonic effective action (\ref{Leffbos}) can be
supersymmetrized using
the superfield technique developped in Ref. \cite{Ivanov}.
The explicit expression in components was written in
Ref. \cite{nonab}:
\begin{eqnarray}
\frac{g^2}{L_1\, L_2 \, L_3} \, {\cal L} \ =\
\frac{1}{2 \, f^2} \, \dot{C}^j\, \dot{C}^j +
\frac{\rm i}{2 f^2} \, \left(
\bar{\Psi}\dot{\Psi} -   \dot{\bar{\Psi}}{\Psi}
\right)
- \frac{\partial_i f}{f^3} \epsilon^{ijk} \, \dot{C}_j \bar{\Psi}
\sigma_k \Psi + \nonumber \\
+ \frac{D^2}{2 \, f^2} + \frac{D \, \partial_i f}{f^3}
\, \bar{\Psi} \sigma_i \Psi - \frac{1}{8} \partial^2
 \left(\frac{1}{f^2}\right)
 \left(\bar{\Psi}\right)^2 \left(\Psi\right)^2\ ,
\label{SQM1}
\end{eqnarray}
where $\Psi = \psi f$, $\psi$ and $\bar\psi$ being
 the canonically conjugated variables
of Eq. (\ref{Heff});
 $D$ is the auxilary field.
The action corresponding to the
Lagrangian (\ref{SQM1}) is invariant under the transformations
   \be
\delta_\epsilon \, C_k &=& \bar \epsilon \sigma_k \Psi
+ \bar {\Psi} \sigma_k \epsilon \ ,\nonumber \\
\delta_\epsilon \Psi_\alpha  &=&  -{\rm i}\, (\sigma_k
\epsilon)_\alpha \dot C_k
+ \epsilon_\alpha  D\ , \nonumber \\
\delta_\epsilon  D &=&
{\rm i}\left(\dot{\bar{\Psi}}  \epsilon
-  \bar\epsilon  \dot{\Psi} \right)  \ .
 \label{SUSY}
  \ee

The chopoff procedure works also for nonsupersymmetric theories.
Consider the simplest case of scalar QED.
The one--loop correction to the effective Lagrangian is just
 \be
\label{dLscal}
\delta {\cal L}_{\rm scalar\ QED} \ =\
-{\rm i} \log \det (-\partial_0^2 - A_k^2) \ .
 \ee

\begin{figure}
     \begin{center}
        \epsfxsize=150pt
        \epsfysize=50pt
        \vspace{-5mm}
        \parbox{\epsfxsize}{\epsffile{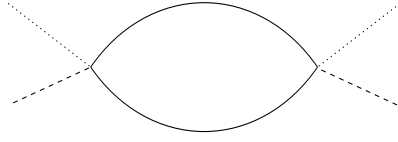}}
        \vspace{5mm}
    \end{center}
\caption{Renormalization of the kinetic term in 
scalar QED in the QM limit. The dashed lines correspond to $\dot{\vec{A}}$
and the pointed lines --- to $\vec{A}$. }
\label{scalQEDfig}
\end{figure}

The double derivative term is given by the graph depicted in
Fig. 2. We obtain
 \be
\label{dLres}
\delta {\cal L}_{\rm scalar\ QED}  =
-\left.(\vec{A} {\dot \vec{A}})^2 \frac {\partial^2}{\partial
\epsilon^2}
\int_{-\infty}^\infty \frac {d\omega}{2\pi} \frac 1{[\omega^2 +
\vec{A}^2 ]
[(\omega+ \epsilon)^2 + \vec{A}^2]}\right|_{\epsilon = 0}
\nonumber \\
 =\   \frac {(\vec{A} {\dot \vec{A}})^2}{8|\vec{A}|^5} \ .
  \ee
Restoring the contribution of the higher Fourier modes,
$\vec{A} \to \vec{A} - 2\pi \vec{n}/L$, and performing the
summation over
$\vec{n}$, with averaging over directions $n_j n_k \equiv
\frac 13 \vec{n}^2\delta_{jk}$, we reproduce the known result
for the one--loop $\beta$ function in the scalar QED,
  \be
\label{chargscalQED}
\frac 1{e^2(L)}\left|_{\rm scalar \ QED} \ =\ \frac 1{e_0^2} +
\frac {1}{24\pi^2}
\ln(\Lambda_{UV}L)  \right.\ .
 \ee
Therefore, the chopoff method works also for
nonsupersymmetric theories. Actually, it  was applied
 before to pure Yang--Mills theory in Ref. \cite{Lusch}.
There are three important distinctions, however:
\begin{itemize}
\item In the nonsupersymmetric case, 
the effective Lagrangian  calculated in the
Abelian background (\ref{decompos})
involves besides the kinetic term
also the potential part $\propto |\vec{C}|$. As a result, in the
non--Abelian case, the true slow variables are not only Abelian, but
all zero Fourier modes of the gauge potential, and the
Born--Oppenheimer Lagrangian becomes more complicated.
 \item A  remark related to the previous one  is that the
metric in Eq. (\ref{dLres}) is not conformally flat as it is in
Eq. (\ref{Leffbos}).
 \item
The graph in Fig. 2
which determines the correction to the effective Lagrangian in the
QM limit is quite different from the standard graph giving the
$\beta$
function in four dimensions. In particular, the former does
not diverge in the ultraviolet in four dimensions.
  \end{itemize}

\section{Nonrenormalisation theorems.}

In the dimensionally reduced  ${\cal N} = 4$
SYM theory (alias, maximally supersymmetric quantum mechanics, alias
matrix model, alias the system of $D0$-branes), the corrections
to the metric on the moduli space are absent. There are $D$-brane
arguments in favor of this conclusion \cite{Duglas}, it was
confirmed by explicit calculation  \cite{sestry}, and
finaly {\it proven} using simple symmetry arguments \cite{Sethi}.
To make the paper self-contained, we present here a somewhat
refined version of these arguments.

In the maximally supersymmetric $SU(2)$ theory the effective
Lagrangian is written in terms of a 9--dimensional vector
$C_k$ and a {\it real} 16--component $SO(9)$ spinor
$\lambda_\alpha$. The Lagrangian must be invariant with respect to
the supersymmetry transformations
   \be
  \label{SUSY9}
\delta_\epsilon C_k &=& -{\rm i} \, \epsilon \gamma_k \lambda\ ,
\nonumber \\
\delta_\epsilon \lambda_\alpha &=&
\left(\gamma_k \dot{C}_k \epsilon \right)_\alpha
+ \left[ M(c) \epsilon \right]_\alpha\ ,
 \ee
where $\epsilon$ is a real Grassmann spinor and $\gamma_k$ are
the 9--dimensional $\gamma$ matrices, $\gamma_j \gamma_k +
\gamma_k \gamma_j = 2\delta_{jk}$. They are all real and symmetric.
The transformations (\ref{SUSY9}) represent an analog of
(\ref{SUSY})
with the auxiliary field expressed out.

The commutator of two SUSY transformations with parameters
$\epsilon_1$ and $\epsilon_2$ should amount to a time translation.
A trivial calculation gives
   \be
  \label{closure}
[\delta_1, \delta_2] C_k \ =\ -2\, {\rm i} \,
\epsilon_2 \epsilon_1 \dot{C}_k
- {\rm i} \, \epsilon_2 \{\gamma_k M + M^T \gamma_k \} \epsilon_1\ ,
  \ee
and we conclude that
 \be
\label{gamMT}
\gamma_k M + M^T \gamma_k = 0\ .
 \ee
As was noticed in Ref.\cite{Sethi},  this implies that $M = 0$.
Let us prove it.
On the first step, note that any $M$ satisfying (\ref{gamMT})
commutes with all generators $J_{kj} =
\frac 14[\gamma_k, \gamma_j]$ of $Spin(9)$. This means that, for
any set $\lambda$ belonging to the spinor
representation of $Spin(9)$, the set
$M\lambda$  also forms a spinor representation. Hence
$M = \xi R$, where $\xi$ is a real number and
$R \in Spin(9)$. But $R$ commutes with all generators of $Spin(9)$
and should belong to the center of $Spin(9)$, i.e. $M = \pm \xi I$.
Then (\ref{gamMT}) tells us that $\xi = 0$.

\vspace{.2mm}

When proving this, we used implicitly the fact that the real
spinor representation of $SO(9)$ is {\it irreducible}. If
it could be decomposed in a direct sum of two other
representations, we could choose $M$ as
diag$(\xi_1 z_1, \ \xi_2 z_2)$, with
$z_1$ and $z_2$ belonging to the center of the group
in the corresponding subspaces. Such $M$ would not be necessarily
proportional to $I$.
This discussion is not purely academic. Actually, for $Spin(3)$
and $Spin(5)$, where real spinor representations are
 reducible, nontrivial matrices satisfying (\ref{gamMT})
{\it exist}.

The only structure not involving higher derivatives\footnote{Higher
derivative terms, in particular the term
$\propto (\dot{C}_k \dot{C}_k)^2$ and its superpartners are
allowed. See \cite{sestry,Plefka} for detailed discussion.}
 and
 invariant with respect to the transformations
(\ref{SUSY9}) with $M=0$ is
$$\frac 12 \left[ \dot{C}_k^2 + {\rm i} \lambda \dot{\lambda} \right]
\ .$$
Nontrivial corrections to the metric are not allowed.
Bearing in mind the  discussion in the previous section, this
simultaneously proves that the $\beta$ function in ${\cal N} = 4$
SYM theory vanishes exactly in all loops.

In the   ${\cal N} = 2$ case, the corrections to the metric survive,
but the presence of 4 different complex supercharges
 dictates  that  the function
$f^{-2}(\vec{C})$  ($\vec{C}$ is now a 5--dimensional  vector.
In four dimensions this corresponds  to the gauge potential and a complex
 scalar.)
is not arbitrary, but should be a harmonic function,
$\Delta^{(5)} f^{-2}(\vec{C}) = 0$ \cite{rumyn,Zupnik}.
The $O(5)$ invariance, which is
manifest in the chopoff quantum--mechanical limit, tells us then
that
the only allowed form of the effective Lagrangian is
 \be
\label{fN=2}
{\cal L}^{\rm eff} = \frac {\dot{\vec{C}}^2}2 \left( 1 +
\frac {\rm const} {|\vec{C}|^3} \right)\ ,
  \ee
i.e. all the corrections beyond one loop vanish.
But this means
also that multiloop corrections to the $\beta$ function vanish
in this case.

\section{Effective action in (1+1) dimensions.}

Consider the gauge SYM theory with $SU(2)$ gauge group compactified
on $T^2$ rather than on $T^3$. The
low--energy dynamics is described by an effective  (1+1) dimensional
field theory depending on {\it two} bosonic variables $C_{1,2}(z,t)$
and their fermionic superpartners. The effective Lagrangian
represents in this
case a supersymmetric K\"ahlerian $\sigma$-model \cite{folklore}.
The  corresponding K\"ahlerian manifold represents a 2--torus dual
to the
spatial
torus. The precise form of the metric can be determined by a simple
one--loop calculation.
It can be done along the same lines as in Sect. 2.
Calculating the
graphs in Fig. 1, we obtain for the correction to
the effective Lagrangian density
 \be
 \label{cormet11}
\Delta {\cal L}_{\rm bos}^{d=2} \ =\
-3 (\partial_\alpha C_j)^2 \int_{-\infty}^\infty \frac{d^2p}{(2\pi)^2}
\frac 1{(p^2 + C_j^2)^2}
\ = \  \frac {(\partial_\alpha {C}_j)^2}{4\pi C_j^2}  \ ,
 \ee
where $\alpha = 0,3$ and $j = 1,2$. We obtain
 \be
\label{Leffbos11}
\frac {g^2}{L^2} {\cal L}^{d=2}_{\rm bos}\ = \
\frac 12 (\partial_\alpha C_j)^2 {\tilde f}^{-2}(C_j)\ ,
 \ee
where
 \be
\label{ftild}
{\tilde f}(C_j) \ =\
 1 + \frac {3g^2}{4\pi L^2 C_j^2} \ .
\ee
The fermion terms can be restored by supersymmetrizing. The full
effective Lagrangian is
 \be
\label{Kahler}
  g^2 {\cal L}^{d=2} \ =\
\int d^2\theta d^2 \bar\theta
\left[ \bar\Phi \Phi - \frac 3{4\pi} \ln \Phi \ln \bar\Phi \right]
\ ,
 \ee
where $\Phi$ is a chiral superfield with the lowest component
$\phi =  {L(C_1 + {\rm i} C_2)}/\sqrt{2} $.
 When deriving (\ref{Leffbos11}), we neglected higher Fourier
modes associated with compactified directions. This is
justified if $|\phi| \ll 1$. (On the other hand, we have
to keep $|\phi| \gg g$, otherwise higher loop corrections
become relevant.) For $|\phi| \sim 1$, the higher Fourier
modes should be
taken  into
account, which can be easily done
 by substituting $C_j \to C_j - 2\pi n_j/L$ and
performing the sum over integer 2--dimensional $n_j$.

It is very instructive to explore the relationship between the
effective $d=2$ Lagrangian (\ref{Kahler}) and our nonstandard
(0+1)--dimensional
$\sigma$ model (\ref{SQM1}).
To begin with, let us study the geometric structure of the
Lagrangian (\ref{SQM1}). To this end, we
integrate out the auxiliary field $D$ and
express the Lagrangian in the form
  \be
\frac{g^2}{L_1\, L_2 \, L_3} \, {\cal L} \ =\
\frac{1}{2} g_{jk} \, \dot{C}^j\, \dot{C}^k +
\frac{\rm i}{2} \, \left(
\bar{\psi}\dot{\psi} -   \dot{\bar{\psi}}{\psi}
\right)
+ {\rm i} \, \omega_j^{ab}  \, \dot{C}^j \bar{\psi}
\sigma^{ab} \psi + \nonumber \\ +
 \frac{1}{4} \left[ f (\partial^2 f) - 2
 (\partial_j f)^2 \right]
 \left(\bar{\psi}\right)^2 \left(\psi\right)^2\ ,
\label{LintD}
    \ee
where we have raised the index of the vector
$C^j$ indicating its contravariant
nature, $g_{jk} = f^{-2} \delta_{jk}$, $\psi = \Psi/f$,
$\sigma^{ab} = \frac i2 \epsilon^{abc} \sigma^c$ is the
generator of rotations in the tangent space, and
  \be
\label{omega2}
\omega_i^{ab} = \delta^b_i \, \partial^a \log{(f)} -
\delta^a_i \, \partial^b \log{(f)}
   \ee
is the spin connection on a conformally flat
manifold with the natural choice of the  dreibein,
$e_j^a = f^{-1}\delta_j^a$.

 We have succeeded in presenting the
bifermion term in a nice geometric form. However, the 4--fermion
term in Eq. (\ref{LintD}) does not have an
obvious geometric interpretation. In particular, its coefficient
{\it is not} a 3--dimensional scalar
curvature.

To establish the relation of (\ref{LintD}) to the
K\"ahlerian model, let us consider the original theory
on an {\it asymmetric} torus $L_3 \gg L_1 = L_2 \equiv L$. The
effective QM model is given by the Lagrangian (\ref{Leffbos11})
with $f(C_{1,2}, C_3)$ written in Eq. (\ref{fasym}). If $L_3$ is very
large, the range where $C_3$ changes is very small. In the limit
$L_3 \to \infty$, $C_3$ is frozen to zero, but to perform this
limit, we cannot just set $C_3 = 0$ in Eq. (\ref{fasym}),
but rather
{\it average} over $C_3$ within the range
$0 \leq C_3 \leq 2\pi/L_3$ and simultaneously perform the summation
over $n_3$. This amounts to calculating the integral
   $$\frac 1{f^2(C_3, C_{1,2})} \to 1 -
\frac {3g^2}2  \sum_{n_j} \int \frac
{da}{2\pi}\ \frac 1{[\sum_{j=1,2} (LC_j - 2\pi n_j)^2 +
a^2]^{3/2}}\ ,$$
    which exactly gives the correction to the metric in Eq.
(\ref{Leffbos11}), (\ref{ftild}).

The Lagrangian (\ref{LintD}) involves also the kinetic term
for the field $C_3$ and the term where $\dot C_3$
multiplies a bifermion structure.
Performing the  functional integral of exp$\{iS\}$
over $\prod_t dC_3(t)$,
we arrive at the expression

\be
   {\cal L} = \ \frac 12 \tilde{g}_{jk} \dot{y}^j  \dot{y}^k  
+ {\rm i}\,\bar{\chi}^a  \,
\left(\delta^{ab} \, \partial_t -  \,
 \omega_j^{ab} \dot{y}^j \right)\,
\chi^b + \frac18 \, R
\, \left(\bar{\chi}\right)^2 \, \left(\chi\right)^2\ ,
 \label{kahler}
\ee
where $ y^j = L c^j/g$, \ $\tilde{g}_{jk} = \tilde{f}^{-2}
 \delta_{jk}$, \ $\chi$ is related to $\psi$ by a unitary 
transformation, 
$\omega_j^{ab}$ is the two-dimensional spin 
connection (\ref{omega2}), 
and $R = 2[\tilde{f} \partial^2 \tilde{f} -
(\partial_j \tilde{f})^2]$ is the   two-dimensional 
 scalar curvature. The Lagrangian (\ref{kahler}) coincides with
the standard Lagrangian of the supersymmetric $\sigma$ model \cite{sigmod}. 
in the QM limit. The $(1+1)$ effective Lagrangian could be obtained
if taking into account the higher Fourier harmonics $\propto \exp\{izn/L_3\}$
of $C_j(z,t)$ and $\psi(z,t)$ in the background.

\section*{Acknowledgements}
A.V.S. is indebted to A. Losev and N. Nekrasov for illuminating
discussions.  E.T.A. acklowledges warm hospitality extended to
him at the University of Nantes where this work was done. 
His work  was partially
supported by the grants RFFI 01-02-17488 and INTAS-00-390.


\begin{thebibliography}{40}
\bibitem{folklore} This is kind of folklore, but we did not see in the 
literature any explicit calculation.
\end{thebibliography}
\end{document}